\documentclass[aps,prd,showkeys,showpacs,amssymb,twocolumn,
amsfonts,epsf,preprintnumbers,nofootinbib,superscriptaddress]{revtex4}

\usepackage[dvips]{graphicx}
\usepackage{bm,latexsym,amsmath,amssymb,amsfonts,color}

\def\be {\begin{equation}}
\def\ee {\end{equation}}
\def\ba {\begin{eqnarray}}
\def\ea {\end{eqnarray}}

\def\bi {\begin{itemize}}
\def\ei {\end{itemize}}
\newcommand\beq{\begin{eqnarray}}
\newcommand\eeq{\end{eqnarray}}
\newcommand{\bea}{\begin{eqnarray}}
\newcommand{\eea}{\end{eqnarray}}

\usepackage{graphicx}
\usepackage{dcolumn}
\usepackage{bm,morefloats,color}

\usepackage{amssymb,bm}

\def\X5sp{{\rm X}_5}
\def\Y3sp{{\rm Y}_3}
\def\Z3sp{{\rm Z}_3}

\begin{document}

\title{Black hole quasinormal modes in a scalar-tensor theory with field derivative coupling to the Einstein tensor}

\author{Masato Minamitsuji}
\affiliation{Multidisciplinary Center for Astrophysics (CENTRA), Instituto Superior T\'ecnico, Lisbon 1049-001, Portugal.}

\begin{abstract}
We investigate the quasinormal modes of a  test massless, minimally coupled scalar field on a static and spherically symmetric black hole in the scalar-tensor theory with field derivative coupling to the Einstein tensor, which is a part of the Horndeski theory with the shift symmetry. In our solution, the spacetime is asymptotically AdS (anti-de Sitter), where the effective AdS curvature scale is determined solely by the derivative coupling constant. The metric approaches the AdS spacetime in the asymptotic infinity limit and precisely recovers the Schwarzschild-AdS solution in general relativity if the coupling constant is tuned to the inverse of the cosmological constant. We numerically find  the lowest lying quasinormal frequency for the perturbation about a test massless, minimally coupled scalar field. The quasinormal frequency agrees with that of the Schwarzschild-AdS solution for the tuned case. For other parameters, in the large black hole limit, as the metric coincides with that of the Schwarzschild-AdS black hole, the quasinormal frequency almost agrees with that of the Schwarzschild-AdS black hole and is insensitive to the value of the cosmological constant. On the other hand,  for a small back hole the real part of the quasinormal frequency decreases as the absolute value of the cosmological constant increases.
\end{abstract} 

\date{\today}

\maketitle

\section{Introduction}

Possible modifications of Einstein's general relativity
on cosmological scales have been explored
as the alternatives to the Concordance Model of cosmology \cite{mod1}.
A naive modification of general relativity provides ghost degrees of freedom arising from the higher derivative terms
as well as inconsistencies with tests of general relativity.
In order to avoid the appearance of the ghost degrees, the equations of motion should be of the second order.
On the other hand,
in order to pass tests of general relativity,
a realistic modification of gravity should contain a mechanism to suppress scalar interactions on small scales \cite{sc1}.
After the investigations of a number of models,
it has turned out that successful models of the modified gravity
can be rewritten in the form of the most general scalar-tensor theory.
Such a theory was already proposed by Hordenski \cite{hor1} forty years ago,
and has been reformulated with the growing interest in the various cosmological problems \cite{hor2}.
Despite the existence of the various derivative interactions, 
in the Horndeski theory the equations of motion remain of the second order \cite{hor1,hor2}.
So far, the Horndeski theory has been mainly applied to the cosmological problems.

There has also been the growing interest in the stationary solutions in this theory.
So far, for the particular class of the Horndeski theory with the shift symmetry, 
the static, spherically symmetric black hole solutions have been derived in Refs. \cite{rinaldi,ana0,ana,min}. 
These works have considered the theory where the first order derivative of the scalar field 
is nonminimally coupled to the Einstein tensor
\bea
\label{action}
S&=&\frac{1}{2}
\int d^4x\sqrt{-g}
\Big[
m_p^2
\big(R-2\Lambda\big)
\nonumber\\
&-&
\big(g^{\mu\nu}-
\frac{z}{m_p^2} G^{\mu\nu}\big)
\partial_\mu \phi
\partial_\nu \phi
\Big],
\eea
where $g_{\mu\nu}$ is the metric,
$g={\rm det}(g_{\mu\nu})$,
$R$ and $G_{\mu\nu}$
are the Ricci scalar and the Einstein tensor associated with the metric $g_{\mu\nu}$, respectively,
the dimensionless parameter $z$ characterizes the strength of the field derivative coupling to the Einstein tensor, 
$\Lambda$ is the cosmological constant,
and $m_p$ is the reduced Planck mass.
Clearly, the theory \eqref{action} is invariant under the shift transformation $\phi\to \phi+{\rm const}$.
Under the assumption that the scalar field is a $C^0$ function of the radial coordinate,
whose derivatives diverge at the horizon
\footnote{The coordinate invariants constructed from the scalar field are regular in the whole spacetime
including the horizon.},
the static black hole solutions have been obtained in \cite{rinaldi,ana0,ana,min,lifshitz}.
These solutions are asymptotically AdS  (anti-de Sitter) \cite{rinaldi,ana0,ana,min}
and asymptotically Lifshitz \cite{lifshitz}, respectively
\footnote{See also \cite{nohair} for a no-hair argument for an asymptotically flat solution 
in this class of scalar-tensor theory without cosmological constant.}.
On the other hand,
in \cite{ana0}, 
it was found that in case that there is the linear time dependence to the scalar field,
the theory \eqref{action} contains the stealth black hole solution where 
the scalar field does not backreact on the metric
and hence the spacetime remains that of Schwarzschild black hole in general relativity
(see also \cite{lifshitz}). 
More recently,
these solutions have been extended to the case of the more general classes of the Horndeski theory \cite{scale,kt}.

The purpose of this paper is to investigate the quasinormal modes of a test massless, minimally coupled scalar field
in the black hole background in the theory \eqref{action}.
The black hole solutions found in the theory \eqref{action} are asymptotically AdS.
Thus properties of the quasinormal modes in our model would have much overlap
with those of the asymptotically AdS black hole solutions found previously.
The first study of the quasinormal ringing for conformally coupled scalar waves in AdS was performed by Chan and Mann \cite{cm}. 　
Horowitz and Hubeny \cite{hh} have investigated quasinormal frequency of a test massless, minimally coupled scalar field on the background of 
the Schwarzschild-AdS black hole. 
For a large AdS black hole, 
both the real and imaginary parts of the quasinormal frequency scale linearly with the black hole temperature,
but for a small AdS black hole they deviate from the proportionality relation. 
The analyses about the Schwarzschild-AdS black hole, including a small AdS black hole,
have been followed by Refs. \cite{sch1,sch2}.
Then the analysis of quasinormal modes in the asymptotically AdS black hole solutions 
have been extended to the various cases:
AdS black holes with other kinds of horizon topology \cite{topo1},
perturbations about the other fields \cite{apa0},
Reissner-Nortstr\"om-AdS black holes \cite{rn1},
Kerr(-Newman)-AdS black holes \cite{kerr1}
and BTZ (Ba\~{n}ados-Teitelboim-Zanelli) black holes \cite{btz0}
(for more recent studies, see \cite{bcs} and references therein).
The search of the quasinormal frequency for the asymptotically AdS black holes has also been activated
by the growing interests in the AdS/CFT (anti-de Sitter/conformal field theory) correspondence  \cite{adscft}.
According to the AdS/CFT correspondence, a black hole in AdS corresponds to a thermal state in the CFT,
and the decay of a test field in the AdS black hole spacetime corresponds to the decay of the perturbed CFT state.
The quasinormal frequencies correspond to the poles of the retarded correlation function in the dual CFT \cite{collapse_ads},
and the imaginary part of the quasinormal frequency determines the relaxation time scale back to the thermal equilibrium \cite{hh}.
Thus the analysis of the quasinormal modes in our model may also provide us a hint for 
the quantum field theory dual to the scalar-tensor theory \eqref{action}
and also a more general class of the Horndeski theory,
which has not been much explored yet.

The construction of this paper is as follows:
In Sec. II, we review the black hole solutions in our model.
In Sec. III, we discuss the general behavior of a massless, minimally coupled scalar field on top of the black hole background.
In Sec. IV, we numerically investigate the quasinormal frequency and discuss their properties.
The last Sec. V is devoted to giving the brief summary and conclusion.

\section{Black hole solutions}

Now, let us review the background black hole solution considered in this paper.
In the case of the positive coupling constant $z>0$,
a static and spherically symmetric black hole solution is given by
\bea
\label{ansatz}
ds^2= -f(r) dt^2+ g(r) dr^2+ r^2 d\Omega^2,\quad \phi=\phi(r),
\eea 
where 
\bea
\label{solp}
f(r)&=&\frac{m_p^3}{3rz(m_p^2-\Lambda z)^2}
\Big\{-24m z
+m_p^3r^3\Big(1-\frac{\Lambda z}{m_p^2}\Big)^2
\nonumber\\
&-&3m_p r z\Big(-1+\frac{\Lambda z}{m_p^2}\Big)\Big(3+\frac{\Lambda z}{m_p^2}\Big)
\nonumber\\
&+&3z^{3/2} \Big(1+\frac{\Lambda z}{m_p^2}\Big)^2{\rm arctan}\big(\frac{m_p r}{\sqrt{z}}\big)
\Big\},
\nonumber\\
g(r)&=&\frac{m_p^4 \big(2z +r^2(m_p^2-\Lambda z)\big)^2}
                {(m_p^2r^2+ z)^2(m_p^2-\Lambda z)^2f(r)},
\nonumber\\
(\phi'(r))^2&=&-\frac{m_p^8 r^2 (m_p^2 +\Lambda z)\big(2 z+r^2 (m_p^2-\Lambda z)\big)^2}
                 {(m_p^2r^2+z)^3(m_p^2-\Lambda z)^2z f(r)},
\eea
the domain of the radial coordinate $r$ is given by $0<r<\infty$,
$d\Omega^2$ represents the line element of a unit two-sphere
and the prime represents the derivative with respect to $r$.
The solution without a cosmological constant $\Lambda=0$ was obtained in \cite{rinaldi},
which was then generalized to the case for a nonzero cosmological constant $\Lambda\neq 0$ in \cite{ana0,ana,min}.
As we mentioned previously,
the asymptotic structure of the spacetime
in the large $r$ limit becomes AdS, irrespective of the sign of the cosmological constant.
This can be seen by taking the large-$r$ limit;
\bea
\label{higo}
f(r)&=&\frac{m_p^2}{3z}r^2+\frac{3m_p^2+\Lambda z}{m_p^2-\Lambda z}+O(r^{-1}),
\nonumber\\
\frac{1}{g(r)}&=&\frac{m_p^2}{3z}r^2+\frac{7m_p^2+\Lambda z}{3m_p^2-3\Lambda z}+O(r^{-1}),
\eea
which is very similar to the static coordinate of the  AdS spacetime.
From the leading order $r^2$ terms,
we can read the effective AdS curvature scale $\ell_{\rm eff}:=\frac{\sqrt{3z}}{m_p}$,
which does not depend on the cosmological constant $\Lambda$.
As $f(r)=0$ has only a single root even for a large positive cosmological constant $\Lambda>0$,
the only event horizon can be formed at the place where $f(r)=0$ and no cosmological horizon exist.
Moreover, choosing a too large positive cosmological constant $\Lambda>\frac{m_p^2}{z}$
gives rise to an additional curvature singularity at the finite  $r=r_s:=\sqrt{\frac{2z}{\Lambda z-m_p^2}}$, 
where $g(r)=0$,
other than the curvature singularity at the center $r=0$.
Thus in order to circumvent the appearance of such a singularity, 
we have to impose  $\Lambda<\frac{m_p^2}{z}$.
Furthermore,
in order to obtain healthy behaviors of the spacetime and the scalar field outside the horizon,
more precisely for the scalar field not to be ghostlike outside the horizon
and also for the black hole to be thermodynamically well-behaved,
for a given coupling constant $z>0$, we also have to impose
\bea
\label{restr}
-2m_p^2\leq  \Lambda z \leq -m_p^2,
\eea
which does not allow for a vanishing or positive cosmological constant \cite{min}.
When the inequality \eqref{restr} is saturated on the upper bound $\Lambda z=-m_p^2$,
the scalar field becomes trivial and the Schwarzschild-AdS solution is recovered:
\bea
f(r)= 1-\frac{2m}{m_p r}-\frac{\Lambda}{3}r^2,
\quad
g(r)=\frac{1}{f(r)}.
\eea
As in \eqref{higo} the leading $r^2$ terms do not vanish unless the limit of $z\to \infty$ is taken \cite{rinaldi},
the solution \eqref{solp} does not contain the asymptotically flat case for the finite coupling constant.
In the rest,  instead of $m$, we parametrize the size of the black hole by the horizon position 
$f(r_h)=0$,
and as for the Schwarzschild-AdS black hole solution,
call the black holes with $\frac{r_h}{\ell_{\rm eff}}\gg 1$ and $\frac{r_h}{\ell_{\rm eff}}\ll 1$
the `large' and `small'  black holes, respectively.

Before proceeding, we explain why we do not consider the case of $z<0$.
For $z<0$, the very similar static solution can be obtained
\cite{ana,min,rinaldi} as
\bea
f(r)&=&\frac{m_p^3}{3rz(m_p^2-\Lambda z)^2}
\Big\{-24m z
+m_p^3r^3\Big(1-\frac{\Lambda z}{m_p^2}\Big)^2
\nonumber\\
&-&3m_p r z\Big(-1+\frac{\Lambda z}{m_p^2}\Big)\Big(3+\frac{\Lambda z}{m_p^2}\Big)
\nonumber\\
&+&3z(-z)^{1/2} \Big(1+\frac{\Lambda z}{m_p^2}\Big)^2{\rm arctanh}\big(\frac{m_p r}{\sqrt{-z}}\big)
\Big\},
\eea
where $g(r)$ and $(\phi'(r))^2$ remain the same as \eqref{solp}. 
The difference from the case of $z>0$ is that 
the radial coordinate has the finite domain $0<r<\frac{\sqrt{-z}}{m_p}$.
In this case, in order to realize $(\phi')^2>0$ outside the horizon,
we have to impose $\Lambda>\frac{m_p^2}{|z|}$.
But at the same time an additional curvature singularity appears at
the finite radius $r=r_s=\sqrt{\frac{2|z|}{m_p^2+\Lambda |z|}} (<\frac{\sqrt{|z|}}{m_p})$
which may not be hidden by the horizon \cite{min}.
For $\Lambda>\frac{m_p^2}{|z|}$, the weak energy condition for the scalar field
is also violated outside the horizon, implying a quantum instability of the solution.
Thus in the rest we will focus on the case $z>0$.

\section{Massless scalar field perturbations}
Having introduced the background solution,
we consider the general behaviors of a test massless, minimally coupled scalar field
obeying the equation of motion $\Box \varphi=0$,
where $\Box$ is the d'Alembertian operator defined in the background given by \eqref{ansatz} with \eqref{solp}.
We have to emphasize that this test scalar field $\varphi$ is different from the scalar field $\phi$ 
in the original scalar-tensor theory \eqref{action}.
Decomposing the test scalar field into the partial modes
\bea
\varphi(x^{\mu})
=\sum_{\ell  m}
      \frac{R(r)}{r} e^{-i\omega t}Y_{\ell m}(\Omega),
\eea
where $R(r)$ is the radial mode function,
$Y_{\ell m}(\Omega)$ is the spherical harmonics on the unit two-sphere S$^2$,
$\ell$ and $m$ are the angular and magnetic quantum numbers,
and $\omega$ is the frequency.
Introducing the tortoise coordinate $dr_\ast=\sqrt{\frac{g}{f}}dr$, 
the radial equation is given by
\bea
\label{radial}
\Big[
-\frac{d^2}{dr_\ast^2}
+V(r)
\Big]
R(r)
=
\omega^2 R(r),
\eea
where the effective potential is given by
\begin{widetext}
\bea
\label{ringo}
V(r)&=&
f(r)
\Big[
\frac{\ell(\ell+1)}{r^2}
+\frac{m_p^2r^2+z}{3m_p^5 z\big(m_p^2 r^3+rz (2-r^2\Lambda)\big)^3}
\Big\{
m_p^{11}r^4 (2r^3+r_h^3)
-3m_p (r-r_h)z^5 \Lambda^2 (-2+3r^2\Lambda)
\nonumber\\
&+&m_p^7 z^2
\big(
25 r^3 +2r_h^3 -16 \Lambda r^5 +6r^7\Lambda^2
+r^2 r_h (9-5r_h^2\Lambda)
+3r^4 r_h\Lambda (-5+r_h^2\Lambda)
\big)
\nonumber\\
&+&m_p^5 z^3
\big(
 6r+18r_h +r^3\Lambda-4r_h^3\Lambda +8r^5\Lambda^2
-2r^7\Lambda^3
+r^4r_h \Lambda^2(3-r_h^2\Lambda)
+r^2 r_h \Lambda (-33+7r_h^2\Lambda)
\big)
\nonumber\\
&+&m_p^9 r^2 z
\big(
8r^3+r_h^3-6r^5\Lambda+r^2(9r_h-3r_h^3\Lambda)
\big)
\nonumber\\
&+&m_p^3 (r-r_h)\Lambda z^4
\big[
12+\Lambda
\big(
-17r^2-2rr_h -2r_h^2+3r^2r_h(r+r_h)\Lambda
\big)
\big]
\nonumber\\
&+&3z^{\frac{3}{2}}
(m_p^2+\Lambda z)^2
\big(m_p^4 r^4+ z^2 (2-3r_h^2\Lambda)+m_p^2 z r^2 (1-r^2\Lambda)\big)
\Big(
 {\rm arctan}\Big(\frac{m_p r_h}{\sqrt{z}}\Big)
- {\rm arctan}\Big(\frac{m_p r}{\sqrt{z}}\Big)
\Big)
\Big\}
\Big].
\eea
\end{widetext}
In the near horizon limit $r\to r_h$, $r_\ast \simeq \ln (r-r_h)\to -\infty$.
On the other hand, in the asymptotic infinity $r\to+\infty$, $r_\ast\to 0-$.
In the asymptotically AdS spacetime,
as in Ref. \cite{hh},
we define the quasinormal modes 
so that they satisfy the ingoing boundary condition at the horizon $r_\ast \to -\infty$
and the regularity condition at the spatial infinity $r_\ast\to 0-$.
As for the case of the Schwarzschild-AdS black hole,
$V(r)$ vanishes at the horizon $r=r_h$, 
because $f(r_h)=0$ and the combination inside the square bracket in Eq. (\ref{ringo})
is regular there.
Outside the horizon, $r>r_h$, $V(r)$ is monotonically increasing as $r$ increases.
As the function of $r_\ast$,
$V(r)$ is exponentially suppressed in the near horizon region 
$r_\ast\to -\infty$ and 
the asymptotic behavior of the mode function is given by 
$R\simeq D_1e^{-i\omega r_\ast} + D_2 e^{i\omega r_\ast}$.
The purely ingoing boundary condition gives $D_2=0$.
On the other hand, in the asymptotic infinity
$R\simeq \frac{E_1}{r^2}+ E_2 r$.
The regularity at the infinity gives $E_2=0$.
As there is no analytic solution of \eqref{radial} with \eqref{ringo},
we need to search the quasinormal frequency numerically.
The details of our method for the numerical search of the quasinormal frequency will be explained later.

When we numerically integrate the equation of motion,
it is convenient to redefine the mode function by
separating the ingoing mode function at the horizon
from the whole radial mode function by $R(r)= e^{-i\omega r_\ast}  \tilde R(r)$,
where $\tilde R(r)$ satisfies the boundary conditions
at the horizon $r=r_h$ and at the infinity $r\to \infty$,
$\tilde R(r_h)=1$ and $\tilde R (\infty)=0$,
respectively.
The radial equation of motion \eqref{radial} then reduces to the form of
\bea
\label{fine}
&&{\tilde R}''
+\Big\{
\Big(\frac{f'}{2f}-\frac{g'}{2g}\Big)
-2i\omega \Big(\frac{g}{f}\Big)^{\frac{1}{2}}
\Big\}
{\tilde R}'
\nonumber\\
&+&
\Big\{
-\frac{\ell (\ell+1)}{r^2}g
+\frac{1}{2r}
\Big(
\frac{g'}{g}
-\frac{f'}{f}
\Big)
\Big\}\tilde R
=0.
\eea
Following \cite{hh},
we firstly confirm that the imaginary part of an eigenvalue $\omega$ is always nonpositive.
Multiplying $\big(\frac{f}{g}\big)^{\frac{1}{2}}$ on \eqref{fine},
we find
\bea
&&\frac{d}{dr}
\Big[
\Big(\frac{f}{g}\Big)^{\frac{1}{2}}
 {\tilde R}'
\Big]
-2i\omega {\tilde R}'
\nonumber\\
&+& 
\Big(\frac{f}{g}\Big)^{\frac{1}{2}}
\Big\{
-\frac{\ell (\ell+1)}{r^2}g
+\frac{1}{2r}
\Big(
\frac{g'}{g}
-\frac{f'}{f}
\Big)
\Big\}\tilde R
=0.
\eea
Then,
multiplying $\tilde R^\ast$ and integrating by parts with the use of the boundary conditions,
$\tilde R(\infty)\to 0$ and $(f/g)|_{r\to r_h}\to 0$,
we find
\bea
\label{junction_model}
&&
\int_{r_h}^{\infty}
dr
\Big\{
\Big(\frac{f}{g}\Big)^{\frac{1}{2}}
\Big| {\tilde R}' \Big|^2
+2i\omega \tilde R^\ast {\tilde R}'
\nonumber\\
&+& 
\Big(\frac{f}{g}\Big)^{\frac{1}{2}}
\Big[
\frac{\ell (\ell+1)}{r^2}g
+\frac{1}{2r}
\Big(
\frac{f'}{f}
-\frac{g'}{g}
\Big)
\Big]
\big|\tilde R\big|^2
\Big\}
=0.
\eea
Both the real and imaginary parts of the above equation must vanish separately.
The imaginary part of \eqref{junction_model} is given by
\bea
0&=& \int_{r_h}^\infty
dr
\Big[
 \omega \tilde R^\ast {\tilde R}'
+ \omega^\ast (\tilde R^\ast)' {\tilde R}
\Big]
\nonumber\\
&=&
\big(\omega-\omega^\ast\big)
\int_{r_h}^{\infty}dr {\tilde R}^\ast {\tilde R}'
-\omega^\ast|\tilde R(r_h)|^2,
\eea
where we have used the regularity at the infinity, $\tilde R(\infty)=0$,
and hence
\bea
\label{ima}
2i  {\rm Im}(\omega)
\int_{r_h}^{\infty}
dr  {\tilde R}^\ast {\tilde R}'
=\omega^\ast 
\big|{\tilde R}(r_h)\big|^2.
\eea
Substituting \eqref{ima} into \eqref{junction_model}, we obtain
\bea
\label{chari}
&&\int_{r_h}^{\infty}
dr
\Big(\frac{f}{g}\Big)^{\frac{1}{2}}
\Big\{
\Big| {\tilde R}' \Big|^2
+\Big[
\frac{\ell (\ell+1)}{r^2}g
+\frac{1}{2r}
\Big(
\frac{f'}{f}
-\frac{g'}{g}
\Big)
\Big]
\big|\tilde R\big|^2
\Big\}
\nonumber\\
&=&-\frac{|\omega|^2}{{\rm Im}(\omega)}\big|{\tilde R}(r_h)\big|^2.
\eea
Since in our background 
$\frac{\ell (\ell+1)}{r^2}g
+\frac{1}{2r}
\Big(
\frac{f'}{f}
-\frac{g'}{g}
\Big)$
is positive and regular outside the horizon $r>r_h$, the left-hand side of \eqref{chari} is positive definite.
Thus in order for the right-hand side to be also positive definite,
we have to impose ${\rm Im}(\omega)<0$,
meaning that the eigenmodes are always decaying.

Following the arguments in \cite{hh},
as for the Schwarzschild-AdS black holes in general relativity,
we also expect the scaling properties of the quasinormal frequency for a large black hole $\frac{r_h}{\ell_{\rm eff}}\gg1$.
Introducing the dimensionless coordinate $y:=\frac{r}{r_h}$ to fix the horizon at $y=1$, 
for the large black hole the metric can be approximated as
\bea
\label{bimeq2}
ds^2&\simeq&
-\Big(-\frac{1}{y}+ y^2 \Big)\frac{r_h^2}{\ell_{\rm eff}^2} dt^2
+\frac{\ell_{\rm eff}^2y}{(-1+y^3)} dy^2
\nonumber\\
&+& r_h^2 y^2 dx_i dx^i,
\eea
Thus the metric in this limit
is invariant under the rescaling of $t\to a t$, $x^i \to a x^i$ and $r_h\to \frac{r_h}{a}$,
where $a$ is a constant.
The time-dependent part of the quasinormal mode function, $e^{-i\omega t}$,
is invariant under the above rescaling,
which results in the rescaling of the frequency as $\omega\to \frac{\omega}{a}$.
Thus, as both $r_h$ and $\omega$ have the same scaling property,
the quasinormal frequency for a large black hole
should scale linearly with the horizon radius, $\omega\propto r_h$.
As argued in \cite{min}, 
the temperature of the black hole is explicitly given by 
$T=\frac{m_p^2(m_p^2r_h^2+z (2-\Lambda r_h^2))}{4\pi r_h z(m_p^2-\Lambda z)}$,
which behaves as that of the Schwarzschild-AdS black hole \cite{hp}.
For a large black hole, the temperature scales as $T=\frac{m_p^2r_h}{4 \pi z}$,
and hence the quasinormal frequency will also linearly scale with the temperature \cite{hh}.
In addition, 
the limiting behavior \eqref{bimeq2} does not depend on the cosmological constant $\Lambda$
and agrees with the large black hole limit of the Schwarzschild-AdS spacetime
with the curvature scale $\ell_{\rm eff}$. 
Thus for a large black hole, the quasinormal frequency should not be sensitive to the value of $\Lambda$.

\section{Quasinormal modes}
We then numerically confirm the quasinormal frequency.
Introducing the new dimensionless coordinate $x:=\frac{r_h}{r}(=y^{-1})$,
the radial equation \eqref{fine} can be rewritten as
\bea
\label{xeq2}
S(x)\frac{d^2}{dx^2}\tilde R
+T(x,\omega)\frac{d}{dx}\tilde R
+U(x)\tilde R=0,
\eea
where 
\bea
\label{expansion}
S(x)&:=&-x^4 \Big(\frac{f}{g}\Big)^{\frac{1}{2}},
\nonumber\\
T(x,\omega)&:=&-x^4
\Big\{
\Big(\frac{f}{g}\Big)^{\frac{1}{2}}
\Big(
\frac{2}{x}
+\frac{f_x}{2f}
-\frac{g_x}{2g}
\Big)
+\frac{2i\omega r_h}{x^2}
\Big\},
\nonumber\\
U(x)&=&
\Big(\frac{f}{g}\Big)^{\frac{1}{2}}
\Big\{
\ell(\ell+1)x^2g
+\frac{x^3}{2}
\Big(
\frac{g_x}{g}
-\frac{f_x}{f}
\Big)
\Big\}.
\eea
There are two regular singular points in Eq. (\ref{xeq2}); 
$x=0$ (the infinity) and $x=1$ (the horizon).
We then derive the boundary conditions.
By taking the limit to the horizon $x\to 1$
with the assumption that $\frac{d^2\tilde R}{dx^2}$ is smooth,
the boundary condition on the horizon is given by
$\tilde R(1)=1$ 
and 
$\frac{d}{dx}\tilde R\Big|_{x=1}= -\frac{U(1)}{T(1,\omega)}$.
On the other hand,  the boundary condition at the asymptotic infinity is given by
the regularity condition $\tilde R(0)=0$.

For the numerical search, we fix $m_p=1$.
Thus the quantities in all the tables and in all the axes of figures in this paper are measured in the unit of $m_p=1$.
Note also that the coupling constant $z$ is dimensionless.
Our method to search the quasinormal frequency is as follows. For a given set of the parameters, we first consider a trial frequency $\omega_1$ and numerically integrate the equation \eqref{xeq2} from the horizon $x=1$ to the spatial infinity $x=0$, with the `initial' conditions at the horizon $\tilde R_1(1)=1$ and $\frac{d}{dx}\tilde R_1\Big|_{x=1}= -\frac{U(1)}{T(1,\omega_1)}$. If $\omega_1$ is not the correct eigenvalue $\omega$, the  numerical solution $\tilde R_1(x)$ fails to satisfy the boundary condition at the infinity, $\tilde R_1(0)\neq 0$.  We then choose another value of the frequency $\omega_2$ close to $\omega_1$, numerically integrate \eqref{xeq2} with the modified `initial' conditions at the horizon $\tilde R_2(1)=1$ and $\frac{d}{dx}\tilde R_2\Big|_{x=1}= -\frac{U(1)}{T(1,\omega_2)}$, and obtain the corresponding numerical solution $\tilde R_2(x)$. Although $\tilde R_2(x)$ would again fail to satisfy the boundary condition for the quasinormal mode at the spatial infinity, $\tilde R_2(0)\neq 0$, the behavior of $\tilde R_2(x)$ around $x=0$ would also be closer to that of the correct quasinormal eigenfunction $\tilde R(x)$, with the correct eigenvalue $\omega$ and $\tilde R(0)=0$, than that of the first one $\tilde R_1(x)$. In this way, we iterate the numerical integration with the new trial frequencies $\omega_3, \omega_4, \omega_5,\cdots$. As the number of iteration $k$ becomes sufficiently large, the behavior of the integrated solution $\tilde R_k(x)$ around $x=0$ is improved very much and the trial value $\omega_k$ can be much closer to the correct quasinormal frequency $\omega$. For each set of the parameters, we iterate the numerical integration until the precision higher than values shown in \cite{hh,sch2} is achieved, as we use values in \cite{hh,sch2} as the reference (see below). 
As the Schwarzschild-AdS solution is recovered for $z=-\frac{1}{\Lambda}$,
we will make use of the case of $z=\frac{1}{3}$ and $\Lambda=-3$ as the reference solution 
which gives the unit effective curvature scale $\ell_{\rm eff}=\frac{\sqrt{3z}}{m_p}=1$ in the unit of $m_p=1$.
Thus as the consistency check, we have confirmed that in the case of $z=\frac{1}{3}$ and $\Lambda=-3$ our method correctly reproduces the quasinormal frequency obtained in Refs. \cite{hh,sch2} both for the large and small Schwarzschild-AdS black holes.

For each horizon size $r_h$ and angular quantum number $\ell$, there is an infinite number of the quasinormal frequencies
labeled by the overtone number $n=0,1,2,\cdots$.
We often order them in terms of the increasing imaginary part of the quasinormal frequency.
The fundamental quasinormal frequency is defind as the one with the lowest imaginary part and labeled by $n=0$. 
Similarly, the first overtone has the second lowest imaginary part and labeled by $n=1$, and so on.
Also, the lowest value of the imaginary part corresponds to the lowest value of the real part,
the second lowest value of the imaginary part corresponds to the second lowest value of the real part,
and so on. 
Thus the increasing overtone number $n$ also corresponds to the increasing energy of the mode.
In this paper, we will present the numerical results for the less damped, lowest lying ($n=0$)
and spherical symmetric ($\ell=0$) modes, for the various choice of the horizon size $r_h$.
We then will comment on the high overtone ($n\geq 1$) and non-spherically symmetric ($\ell\geq 1$) modes,
which are damped more rapidly.

The fundamental quasinormal frequency for $\ell=0$ in the Schwarzschild-AdS spacetime 
has been numerically investigated in \cite{hh} and subsequent works \cite{sch1,sch2}.
Varying either the coupling constant $z$ or the cosmological constant $\Lambda$,
we investigate how the quasinormal frequency $\omega$ scales with these parameters.
Within the bounds \eqref{restr}, we consider the following cases

\vspace{0.2cm}

Case 1): fixing $\Lambda=-3$, $\frac{1}{3}\leq z\leq \frac{2}{3}$,

\vspace{0.1cm}

Case 2): fixing $z=\frac{1}{3}$, $3\leq |\Lambda| \leq 6$.

\vspace{0.2cm}

First, we consider Case 1).
In Case 1), as one increases $z$, 
the effective AdS curvature scale also increases as $\ell_{\rm eff}=\sqrt{3z}$ in the unit of $m_p=1$.
In Table I,  for the various choice of the horizon size $r_h$,
the real and imaginary parts of the quasinormal frequency
$\omega=\omega_r-i\omega_i$ are shown,
where $\omega_r$ and $\omega_i$ are real and positive. 
\begin{table*}
\begin{minipage}[t]{.80\textwidth}
\caption{
Quasinormal frequency $\omega$ for $\Lambda=-3$.
Note that the quantities in the table are measured in the unit of $m_p=1$.
}
\begin{center}
\label{table3}
\begin{tabular}{|c|c|c|c|c|c|c|c|c|
}
\hline
  $r_h$
&\multicolumn{2}{c|}{$0.01$}
&\multicolumn{2}{c|}{$0.03$}
&\multicolumn{2}{c|}{$0.05$}
\\ \hline  
$z$
 & $ \omega_r$ 
 & $ \omega_i$
 & $ \omega_r$ 
 & $ \omega_i$ 
 & $\omega_r$ 
 & $ \omega_i$
\\ \hline
$\frac{1}{3}$
& $2.9738$
& $0.00054711$
& $2.9167$
& $0.0058571$
& $2.8539$
& $0.019329$ 
  \\ \hline
$0.40$
& $2.5423$
& $0.00042496$
& $2.4967$
& $0.0044909$
& $2.4468$
& $0.014671$
 \\ \hline
$0.50$
& $2.0831$
& $0.00031269$
& $2.0488$
& $0.0032583$
& $2.0155$
& $0.010520$
\\ \hline
$0.60$
&  $1.7603$
&  $0.00024380$
&  $1.7331$
&  $0.0025147$
&  $1.7033$
&  $0.0080482$
\\
\hline
$\frac{2}{3}$
& $1.5937$ 
& $0.00021128$
& $1.5699$
& $0.0021677$
& $1.5439$ 
& $0.0069039$
 \\ \hline
\hline
  $r_h$
&\multicolumn{2}{c|}{$0.07$}
&\multicolumn{2}{c|}{$0.10$}
&\multicolumn{2}{c|}{$0.30$}
\\ \hline  \hline
$z$
 & $\omega_r$ 
 & $ \omega_i$
 & $ \omega_r$ 
 & $\omega_i$ 
 & $\omega_r$ 
 & $\omega_i$
\\ \hline
$\frac{1}{3}$
& $2.7882$
& $0.043587$
& $2.6928$
& $0.10096$
& $2.3845$
& $0.70413$ 
  \\ \hline
$0.40$
& $2.3943$
& $0.032959$
& $2.3169$
& $0.076727$
& $2.0532$
& $0.56339$
 \\ \hline
$0.50$
& $1.9715$
& $0.023504$
& $1.9121$
& $0.054899$
& $1.6975$
& $0.42695$
\\ \hline
$0.60$
&  $1.6719$
&  $0.017892$
&  $1.6244$
&  $0.041829$
&  $1.4445$
&  $0.33931$
\\
\hline
$\frac{2}{3}$
& $1.5165$ 
& $0.015305$
& $1.4748$
& $0.035776$
& $1.3136$ 
& $0.29680$
 \\ \hline
\hline
  $r_h$
&\multicolumn{2}{c|}{$0.50$}
&\multicolumn{2}{c|}{$0.70$}
&\multicolumn{2}{c|}{$1.0$}
\\ \hline 
$z$
 & $ \omega_r$ 
 & $ \omega_i$
 & $\omega_r$ 
 & $\omega_i$ 
 & $\omega_r$ 
 & $ \omega_i$
\\ \hline
$\frac{1}{3}$
& $2.3830$
& $1.2972$
& $2.5014$
& $1.8569$
& $2.7982$
& $2.6712$ 
  \\ \hline
$0.40$
& $2.0451$
& $1.0540$
& $2.1386$
& $1.5192$
& $2.3793$
& $2.1981$
 \\ \hline
$0.50$
& $1.6855$
& $0.81377$
& $1.7562$
& $1.1831$
& $1.9437$
& $1.7248$
\\ \hline
$0.60$
& $1.4320$
& $0.65650$
& $1.4888$
& $0.96119$
& $1.6425$
& $1.4101$
\\
\hline
$\frac{2}{3}$
& $1.3008$ 
& $0.57913$
& $1.3510$
& $0.85131$
& $1.4882$ 
& $1.2536$
 \\ \hline
\hline
  $r_h$
&\multicolumn{2}{c|}{$3.0$}
&\multicolumn{2}{c|}{$5.0$}
&\multicolumn{2}{c|}{$7.0$}
\\ \hline  
$z$
 & $ \omega_r$ 
 & $ \omega_i$
 & $ \omega_r$ 
 & $  \omega_i$ 
 & $ \omega_r$ 
 & $ \omega_i$
\\ \hline
$\frac{1}{3}$
&  $5.9158$
&  $8.0012$
&  $9.4711$
& $13.3255$
& $13.1066$
& $18.6517$ 
  \\ \hline
$0.40$
& $4.9527$
& $6.6514$
& $7.9070$
& $11.0941$
& $10.9326$
& $15.5355$
 \\ \hline
$0.50$
& $3.9823$
& $5.2989$
& $6.3384$
& $8.8607$
& $8.7554$
& $12.4177$
\\ \hline
$0.60$
&  $3.3303$
&  $4.3955$
&  $5.2895$
&  $7.3704$
&  $7.3016$
&  $10.3380$
\\
\hline
$\frac{2}{3}$
& $3.0025$ 
& $3.9434$
& $4.7639$
& $6.6246$
& $6.5638$ 
& $9.2978$
 \\ \hline
\hline
  $r_h$
&\multicolumn{2}{c|}{$10$}
&\multicolumn{2}{c|}{$20$}
&\multicolumn{2}{c|}{$50$}
\\ \hline  
$z$
 & $ \omega_r$ 
 & $ \omega_i$
 & $\omega_r$ 
 & $ \omega_i$ 
 & $ \omega_r$ 
 & $ \omega_i$
\\ \hline
$\frac{1}{3}$
&   $18.6070$
&   $26.6418$
&   $37.0449$
&   $53.2787$
&   $92.4937$
& $133.1933$ 
  \\ \hline
$0.40$
&   $15.5132$
&   $22.1961$
&   $30.8745$
&   $44.3962$
&   $77.0795$
&   $110.9940$
 \\ \hline
$0.50$
&   $12.4171$
&   $17.7497$
&   $24.7028$
&   $35.5131$
&   $61.6649$
&   $88.7931$
\\ \hline
$0.60$
&   $10.3514$
&   $14.7841$
&   $20.5876$
&   $29.5906$
&   $51.3882$
&   $73.9928$
\\
\hline
$\frac{2}{3}$
&   $9.3180$ 
&   $13.3010$
&   $18.5297$
&   $26.6292$
&   $46.2497$ 
&   $66.5922$
 \\ \hline
\end{tabular}
\end{center}
\end{minipage}
\end{table*}
\begin{table*}
\begin{minipage}[t]{.80\textwidth}
\caption{
$ a_r(z)$ an $ a_i(z)$ for $\Lambda=-3$.
Note that the quantities in the table are measured in the unit of $m_p=1$.
}
\begin{center}
\label{table3}
\begin{tabular}{|c|c|c|c|c|c|c|c|c|
}
\hline
$z$
 & $ a_r$ 
 & $ a_i$
\\ \hline
$\frac{1}{3}$
& $1.8520$
& $2.6639$
  \\ \hline
$0.40$
& $1.5435$
& $2.2198$
 \\ \hline
$0.50$
& $1.2349$
& $1.7757$
\\ \hline
$0.60$
&  $1.0292$
&  $1.4796$
\\
\hline
$\frac{2}{3}$
&  $0.9263$ 
&  $1.3315$
 \\ \hline
\end{tabular}
\end{center}
\end{minipage}
\end{table*}
In Figs. 1-2, they are plotted as the functions of $r_h$.  
We find that the behavior of the quasinormal frequency is similar to that of the
case of the Schwarzschild-AdS black hole.
As the coupling constant $z$ increases, the values of the quasinormal frequency decrease.
For $z=\frac{1}{3}$, $\omega$ approaches $3$, corresponding to the lowest lying normal mode of the AdS spacetime \cite{bl}.
For a large black hole, we have numerically confirmed that both ${\omega}_r$ and ${\omega}_i$
are proportional to $r_h$.
The proportionality coefficients $\omega_r \approx  a_r (z) r_h$ and $\omega_i \approx  a_i(z) r_h$ are listed in Table II,
and 
\bea
a_r (z)\approx \frac{0.618}{z}\approx \frac{1.85}{\ell_{\rm eff}^2},
\quad
a_i (z)\approx \frac{0.888}{z}\approx  \frac{2.66}{\ell_{\rm eff}^2},
\eea
This almost agrees with the results shown in Ref. \cite{hh}
and is consistent with our previous arguments.
Namely, as the metric in the large black hole limit \eqref{bimeq2} agrees with that of the Schwarzschild-AdS spacetime
with the effective AdS scale $\ell_{\rm ell}$,
$\omega$ for a large black hole also coincides with that of the given Schwarzschild-AdS spacetime.    

\begin{figure*}[ht]
\unitlength=1.1mm
\begin{center}
\begin{picture}(155,50)
  \includegraphics[height=5.5cm,angle=0]{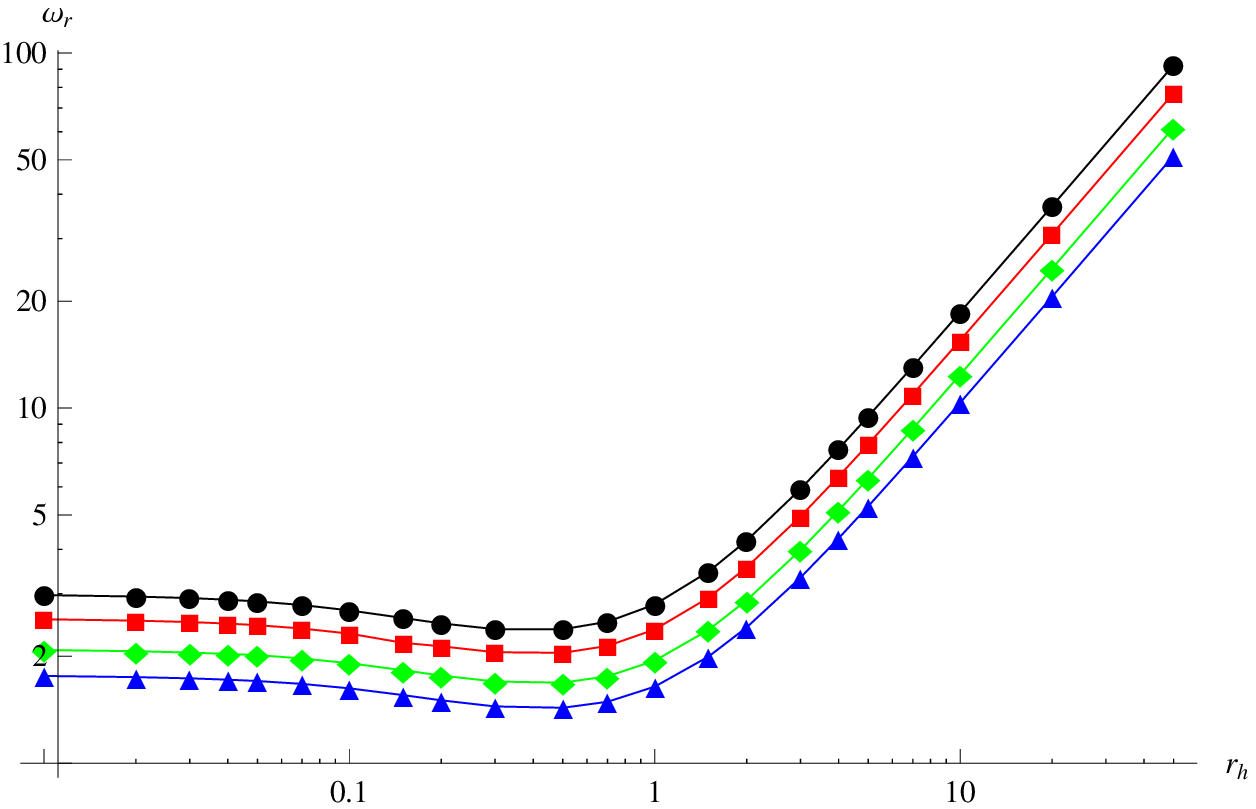}
  \includegraphics[height=5.5cm,angle=0]{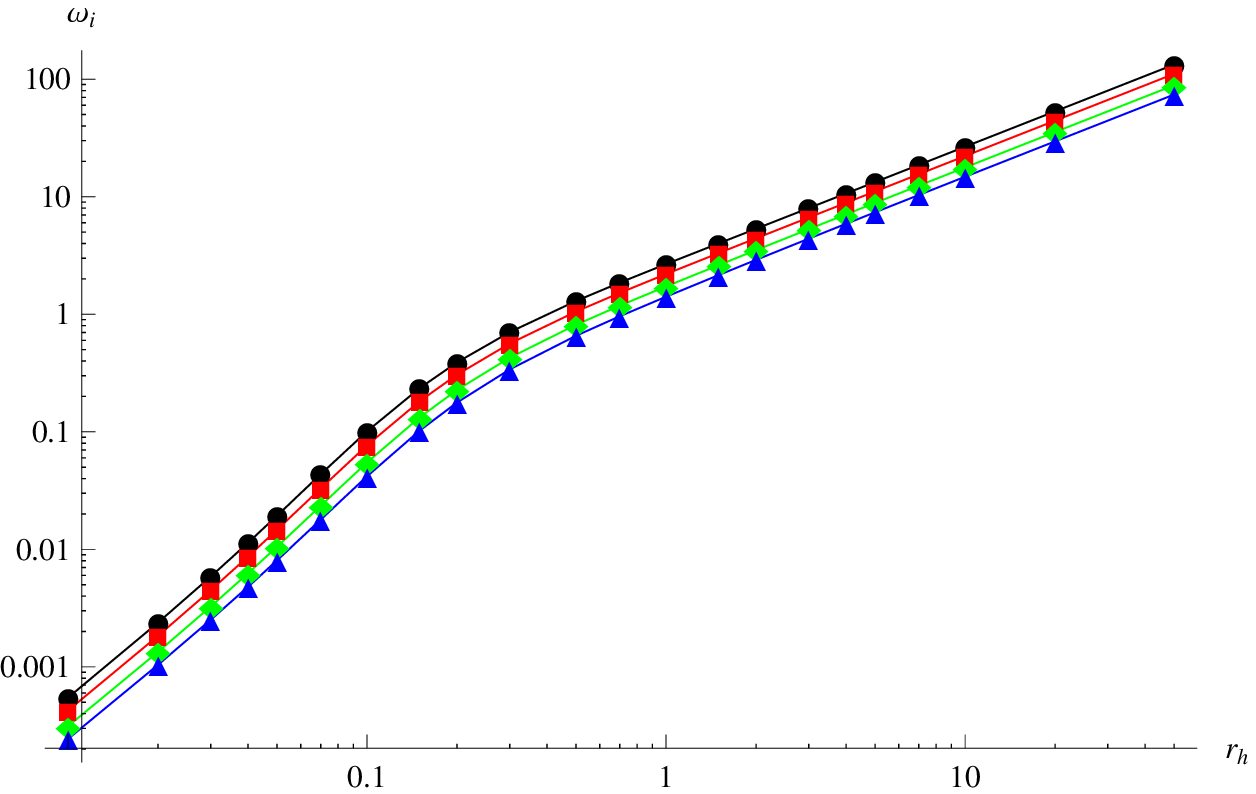}
\end{picture}
\caption{
$\omega_r$ and $\omega_i$ are shown as the functions of $r_h$:
In both panels,
the curves from the top to the bottom
 correspond to $z=\frac{1}{3}$, $0.40$, $0.50$ and $0.60$, respectively.
The case of $z=\frac{1}{3}$ corresponds to that of the Schwarzschild-AdS black hole.
Note that the quantities in the axes are measured in the unit of $m_p=1$.
}
  \label{fig:c1}
\end{center}
\end{figure*} 
\begin{figure*}[ht]
\unitlength=1.1mm
\begin{center}
\begin{picture}(155,50)
  \includegraphics[height=5.5cm,angle=0]{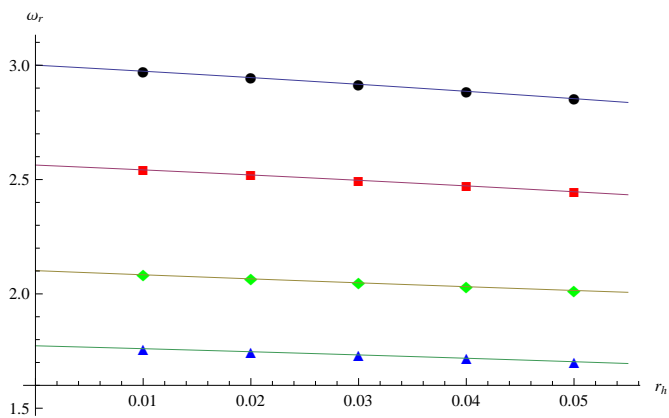}
\end{picture}
\caption{
An enlarged display of $ \omega_r$ as the function of $r_h$ for small black holes:
The curves from the top to the bottom
correspond to $z=\frac{1}{3}$, $0.40$, $0.50$ and $0.60$, respectively.
The case of $z=\frac{1}{3}$, approaching $3.0$ for $r_h\to 0$,
corresponds to that of the Schwarzschild-AdS black hole. 
Note that the quantities in the axes are measured in the unit of $m_p=1$.
}
  \label{fig:c1}
\end{center}
\end{figure*} 

\vspace{0.2cm}
In order to investigate the dependence on the cosmological constant $\Lambda$,
we then consider Case 2).
As in this case $\ell_{\rm eff}=1$ for any value of $\Lambda$,
the deviation from the exact Schwarzschild-AdS black holes
would be able to be made more explicit. 
In Table III, $\omega_r$ and $\omega_i$ are shown for the various choice of the horizon size $r_h$.
\begin{table*}
\begin{minipage}[t]{.80\textwidth}
\caption{
Quasinormal frequency $ \omega$ for $z=\frac{1}{3}$.
Note that the quantities in the table are measured in the unit of $m_p=1$.
}
\begin{center}
\label{table3}
\begin{tabular}{|c|c|c|c|c|c|c|c|c|
}
\hline
  $r_h$
&\multicolumn{2}{c|}{$0.01$}
&\multicolumn{2}{c|}{$0.03$}
&\multicolumn{2}{c|}{$0.05$}
\\ \hline  
$\Lambda$
 & $\omega_r$ 
 & $  \omega_i$
 & $ \omega_r$ 
 & $ \omega_i$ 
 & $\omega_r$ 
 & $ \omega_i$
\\ \hline
$-3$
& $2.9738$
& $0.00054711$
& $2.9167$
& $0.0058571$
& $2.8539$
& $0.019329$ 
  \\ \hline
$-4$
& $2.6713$
& $0.00057213$
& $2.6172$
& $0.0061641$
& $2.5578$
& $0.020393$
 \\ \hline
$-5$
& $2.4360$
& $0.00059377$
& $2.3845$
& $0.0064350$
& $2.3277$
& $0.021330$
\\ \hline
$-6$
&  $2.2471$
&  $0.00061297$
&  $2.1977$
&  $0.0066798$
&  $2.1432$
&  $0.022174$
 \\ \hline
\hline
  $r_h$
&\multicolumn{2}{c|}{$0.07$}
&\multicolumn{2}{c|}{$0.10$}
&\multicolumn{2}{c|}{$0.30$}
\\ \hline 
$\Lambda$
 & $ \omega_r$ 
 & $ \omega_i$
 & $ \omega_r$ 
 & $ \omega_i$ 
 & $ \omega_r$ 
 & $\omega_i$
\\ \hline
$-3$
& $2.7882$
& $0.043587$
& $2.6928$
& $0.10096$
& $2.3845$
& $0.70413$ 
  \\ \hline
$-4$
& $2.4960$
& $0.045814$
& $2.4082$
& $0.10469$
& $2.1521$
& $0.69170$
 \\ \hline
$-5$
& $2.2693$
& $0.047729$
& $2.1879$
& $0.10952$
& $1.9738$
& $0.68031$
\\ \hline
$-6$
&  $2.0876$
&  $0.049411$
&  $2.0118$
&  $0.11020$
&  $1.8323$
&  $0.66999$
 \\ \hline
\hline
  $r_h$
&\multicolumn{2}{c|}{$0.50$}
&\multicolumn{2}{c|}{$0.70$}
&\multicolumn{2}{c|}{$1.0$}
\\ \hline 
$\Lambda$
 & $\omega_r$ 
 & $ \omega_i$
 & $ \omega_r$ 
 & $ \omega_i$ 
 & $\omega_r$ 
 & $ \omega_i$
\\ \hline
$-3$
& $2.3830$
& $1.2972$
& $2.5014$
& $1.8569$
& $2.7982$
& $2.6712$ 
  \\ \hline
$-4$
& $2.1847$
& $1.2662$
& $2.3294$
& $1.8157$
& $2.6570$
& $2.6251$
 \\ \hline
$-5$
& $2.0334$
& $1.2399$
& $2.1991$
& $1.7819$
& $2.5510$
& $2.5884$
\\ \hline
$-6$
& $1.9142$
& $1.2175$
& $2.0969$
& $1.7538$
& $2.4682$
& $2.5584$
 \\ \hline
\hline
  $r_h$
&\multicolumn{2}{c|}{$3.0$}
&\multicolumn{2}{c|}{$5.0$}
&\multicolumn{2}{c|}{$7.0$}
\\ \hline 
$\Lambda$
 & $ \omega_r$ 
 & $ \omega_i$
 & $\omega_r$ 
 & $\omega_i$ 
 & $\omega_r$ 
 & $\omega_i$
\\ \hline
$-3$
&  $5.9158$
&  $8.0012$
&  $9.4711$
& $13.3255$
& $13.1066$
& $18.6517$ 
  \\ \hline
$-4$
&  $5.8579$
&  $7.9730$
&  $9.4355$
&  $13.3075$
&  $13.0811$
&  $18.6385$
 \\ \hline
$-5$
&  $5.8175$
&  $7.9516$
&  $9.4089$
&  $13.2939$
&  $13.0621$
&  $18.6286$
\\ \hline
$-6$
 & $5.7808$
&  $7.9348$
&  $9.3882$
&  $13.2832$
&  $13.0471$
&  $18.6209$
 \\ \hline
\hline
  $r_h$
&\multicolumn{2}{c|}{$10$}
&\multicolumn{2}{c|}{$20$}
&\multicolumn{2}{c|}{$50$}
\\ \hline  
$\Lambda$
 & $ \omega_r$ 
 & $\omega_i$
 & $\omega_r$ 
 & $\omega_i$ 
 & $\omega_r$ 
 & $ \omega_i$
\\ \hline
$-3$
&   $18.6070$
&   $26.6418$
&   $37.0449$
&   $53.2787$
&   $92.4937$
& $133.1933$ 
  \\ \hline
$-4$
&   $18.5891$
&   $26.6324$
&   $37.0359$
&   $53.2740$
&   $98.4900$
&   $133.1910$
 \\ \hline
$-5$
&   $18.5756$
&   $26.6254$
&   $37.0292$
&   $53.2705$
&   $92.4873$
&   $133.1900$
\\ \hline
$-6$
&   $18.5652$
&   $26.6200$
&   $37.0239$
&   $52.2677$
&   $92.4852$
&   $133.1890$
 \\ \hline
\end{tabular}
\end{center}
\end{minipage}
\end{table*}
\begin{table*}
\begin{minipage}[t]{.80\textwidth}
\caption{
${b}_r(\Lambda)$ and ${ b}_i(\Lambda)$ for $z=\frac{1}{3}$.
Note that the quantities in the table are measured in the unit of $m_p=1$.
}
\begin{center}
\label{table3}
\begin{tabular}{|c|c|c|c|c|c|c|c|c|
}
\hline
$\Lambda$
 & $ b_r$ 
 & $ b_i$
\\ \hline
$-3$
& $1.8520$
& $2.6639$
  \\ \hline
$-4$
& $1.8516$
& $2.6637$
 \\ \hline
$-5$
& $1.8513$
& $2.6636$
\\ \hline
$-6$
& $1.8510$
& $2.6634$
 \\ \hline
\end{tabular}
\end{center}
\end{minipage}
\end{table*}

In Figs. 3-4, they are plotted as the functions of $r_h$.  
Both $\omega_r$ and $\omega_i$ decrease as $|\Lambda|$ increases. 
For a large black hole, however,
as all curves are degenerate, the dependence of $\omega$ on $\Lambda$ becomes very week,
which was already expected from our argument below \eqref{bimeq2}.
In addition, 
for a large black hole,
both ${\omega}_r$ and ${\omega}_i$ are proportional to $r_h$.
The proportionality coefficients  
$\omega_r \approx  b_r(\Lambda) r_h$ and $\omega_i \approx   b_i(\Lambda) r_h$ are listed in Table IV,
which are also not sensitive to the value of $\Lambda$.
On the other hand, for a small black hole
$\omega_r$ does not exhibit the degenerate behavior as for a large black hole
and as $|\Lambda|$ increases the quasinormal frequency decreases.
\begin{figure*}[ht]
\unitlength=1.1mm
\begin{center}
\begin{picture}(155,50)
  \includegraphics[height=5.5cm,angle=0]{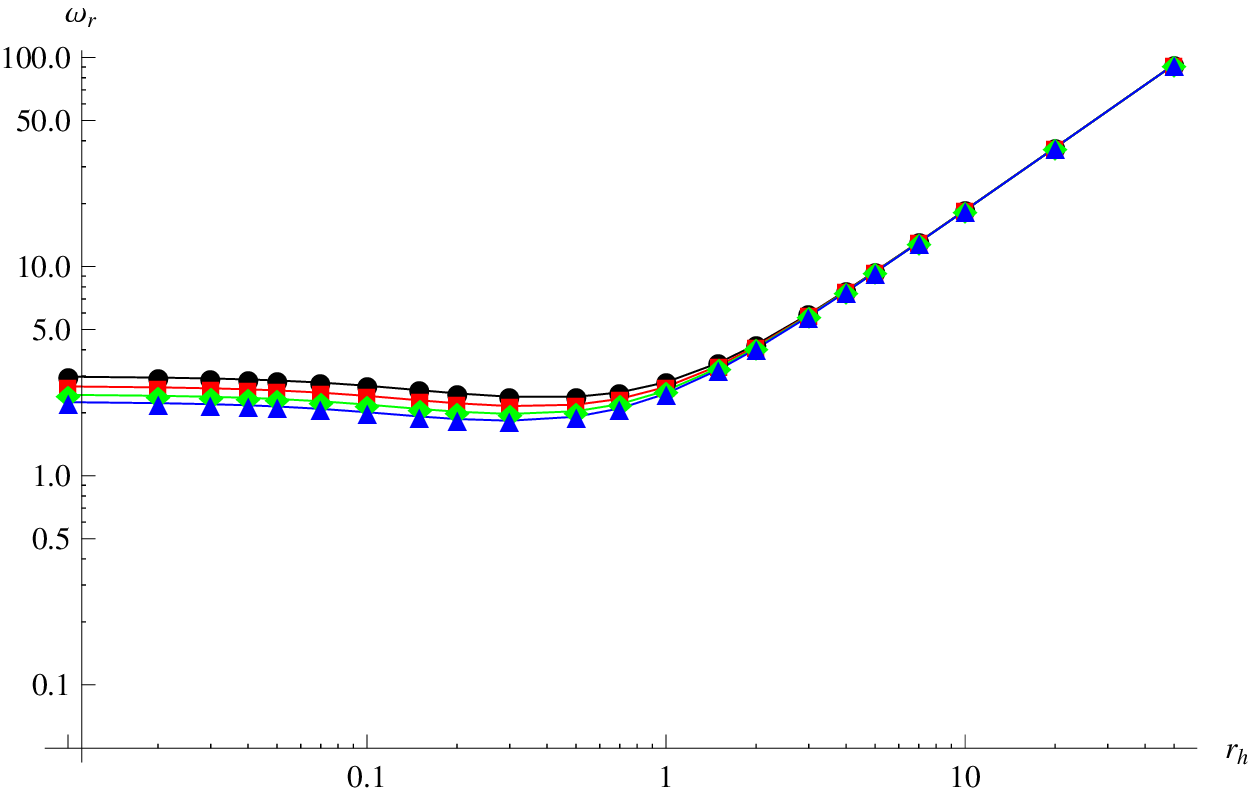}
  \includegraphics[height=5.5cm,angle=0]{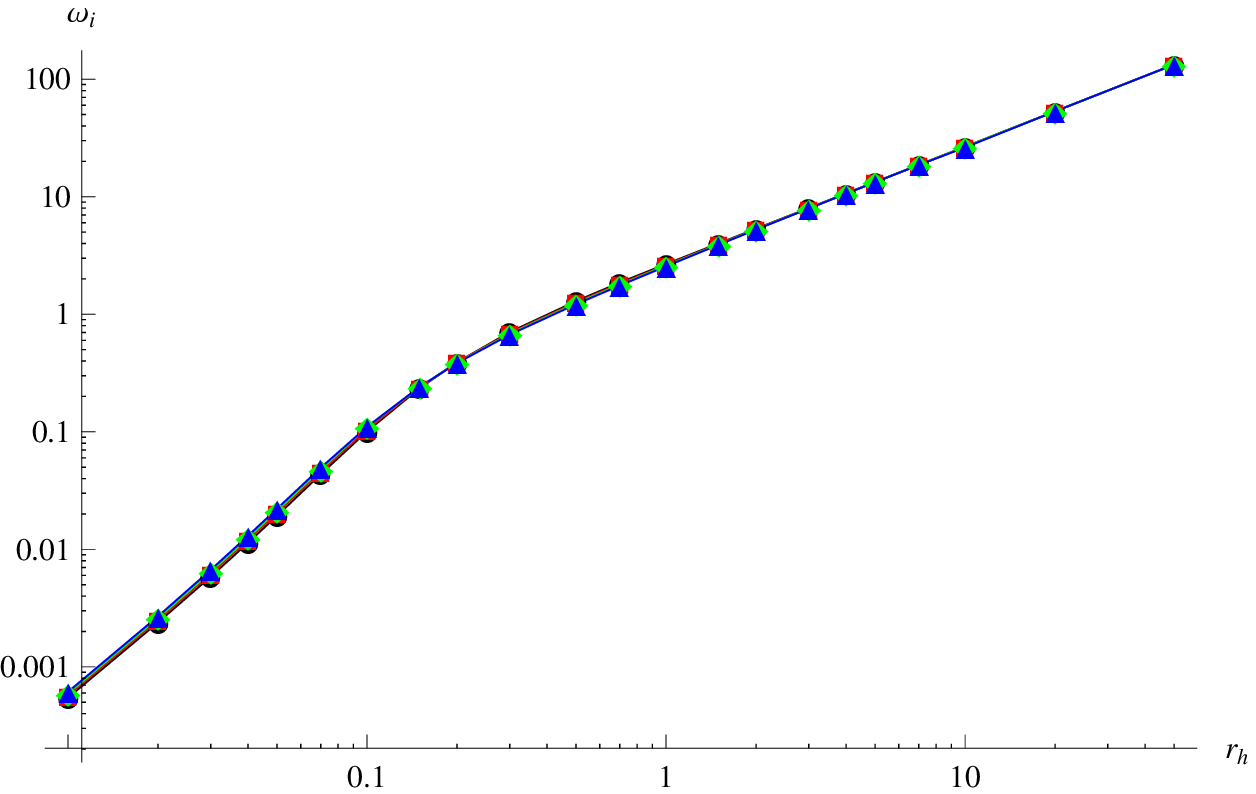}
\end{picture}
\caption{
$\omega_r$ and $\omega_i$ are shown as the functions of $r_h$:
In the panel of $ \omega_r$,
the curves from the top to the bottom correspond to $|\Lambda|=3$, $4$, $5$ and $6$, respectively.
The curves of the same type are also used in the panel for $\omega_i$
but are almost degenerate.
The case of $|\Lambda|=3$ corresponds to that of the Schwarzschild-AdS black hole.
Note that the quantities in the axes are measured in the unit of $m_p=1$.
}
  \label{fig:c1}
\end{center}
\end{figure*} 
\begin{figure*}[ht]
\unitlength=1.1mm
\begin{center}
\begin{picture}(155,50)
  \includegraphics[height=5.5cm,angle=0]{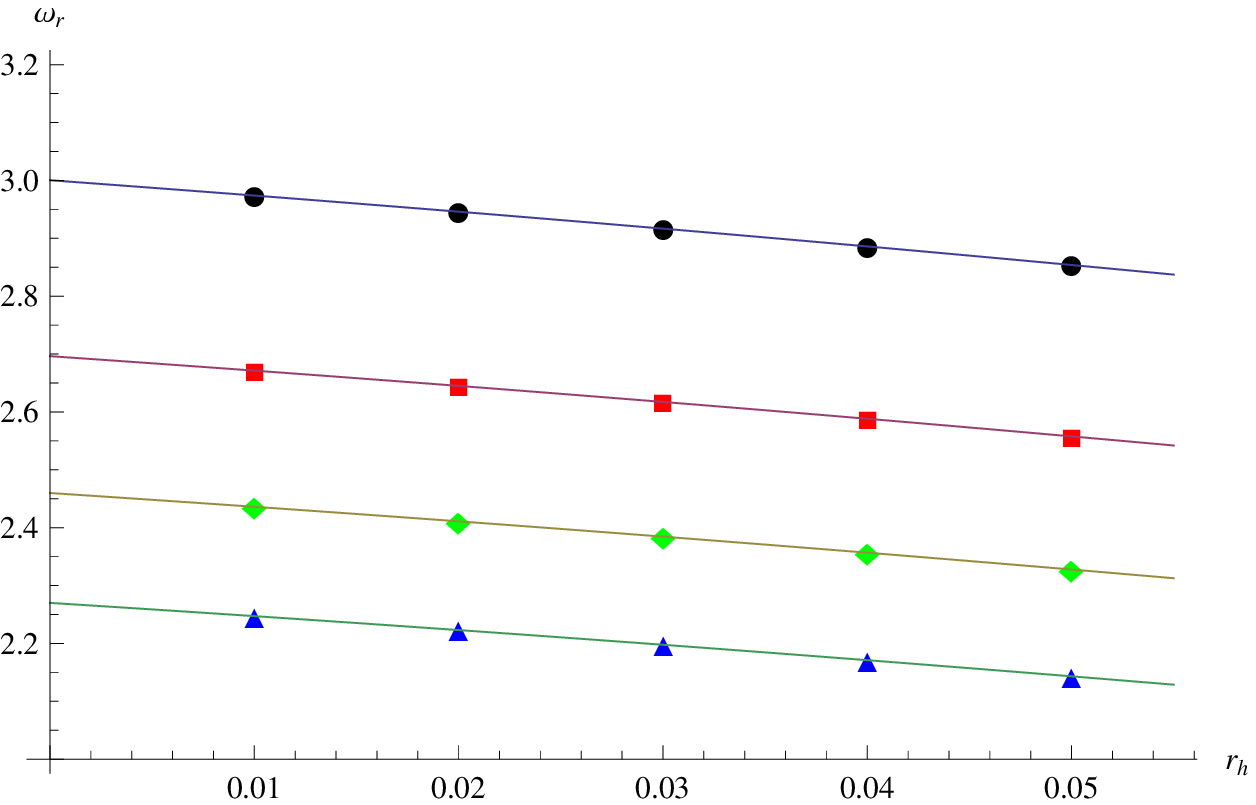}
\end{picture}
\caption{
An enlarged display of ${\omega}_r$ as the function $r_h$ for small black holes:
The curves from the top to the bottom
correspond to $|\Lambda|=3$, $4$, $5$ and $6$, respectively.
Note that the quantities in the axes are measured in the unit of $m_p=1$.
}
  \label{fig:c1}
\end{center}
\end{figure*} 

We then comment on the high overtone ($n\geq 1$) and non-spherically symmetric ($\ell\geq 1$) modes.
In the case of the Schwarzschild-AdS spacetime,
a complete search of the quasinormal frequency including these highly damped modes
was performed in \cite{hightone}.
As argued below \eqref{bimeq2}, in the large black hole limit
our black hole spacetime resembles the Schwarzschild-AdS spacetime with $\ell_{\rm eff}$
solely determined by the coupling constant $z$. 
Hence, we expect that especially for a large black hole
the quasinormal frequency is almost identical to that of the Schwarzschild-AdS spacetime with $\ell_{\rm eff}$
obtained in \cite{hightone},  
and also insensitive to the value of $\Lambda$.
From \cite{hightone},
we expect
for the $\ell=0$ modes, 
in the large black hole and high overtone limits $(r_h,n)\to \infty$
the quasinormal frequency will behave as 
\bea
\frac{\omega \ell_{\rm eff}^2}{r_h}
\approx (1.299-2.25i)n+1.856-2.675i,
\eea
which is evenly spaced.
In addition, from \cite{hightone}
the asymptotic behavior of the spacing of $\omega$, $(1.299-2.25i)\frac{r_h}{\ell_{\rm eff}^2}$,
is expected to hold also for the any other value of the angular quantum number $\ell$,
although the offset value for $n=0$ is $\ell$-dependent.  
On the other hand, for a small black hole 
we expect more sensitivity to the value of $\Lambda$
than that for a large black hole,
but also an evenly-spaced asymptotic behavior of the quasinormal frequencies in the large $n$ limit,
as for the case of the Schwarzschild-AdS black hole \cite{hightone}.

\section{Conclusions}
Before closing this paper, we will give the brief summary and future prospects.
We have numerically investigated the lowest lying quasinormal modes of a  test massless, minimally coupled scalar field in a static and spherically black hole background
in a class of the Horndeski scalar-tensor theory with field derivative coupling to the Einstein tensor \eqref{action},
obtained in Refs. \cite{rinaldi,ana0,ana,min}.
We have considered the perturbation about a test massless, minimally coupled scalar field.
As our black hole spacetime is asymptotically AdS with the curvature scale $\ell_{\rm eff}=\frac{\sqrt{3z}}{m_p}$,
the properties are very similar to those of the Schwarzschild-AdS black holes in general relativity.
The metric reproduces that of the Schwarzschild-AdS black hole solution for $\Lambda z=-m_p^2$
where the scalar field becomes trivial.
In order to make the solution healthy outside the horizon, 
we have focused on the range \eqref{restr} for a fixed coupling constant.

For a fixed $\Lambda$, as the coupling constant $z$ increases,
both the real and imaginary parts of the quasinormal frequency decrease and scale as $\ell_{\rm eff}^{-2}$.
For a fixed coupling constant $z$,
as the metric in the large black hole limit agrees with that of the Schwarzschild-AdS black hole with
a fixed $\ell_{\rm eff}$ which does not depend on $\Lambda$,
the quasinormal frequency in this limit is not sensitive to $\Lambda$. 
For a small black hole,
as the metric deviates from that of the Schwarzschild-AdS spacetime
and the dependence of the metric on $\Lambda$ becomes more explicit,
the quasinormal frequency also becomes more sensitive to $\Lambda$ compared to the case of a large black hole
and as $|\Lambda|$ increases $\omega_r$ decreases.
There would be various extensions of the present analyses,
e.g., to include the effects of the mass and couplings of the test scalar field
and to investigate the perturbation by another test field with a different spin,
especially about the gravitational field,
all which may provide us further hints to distinguish two theories more explicitly.

\section*{Acknowledgement}
This work was supported by the FCT-Portugal through Grant No. SFRH/BPD/88299/2012.



\end{document}